\documentclass[preprint]{iucr}              

 \journalcode{A}              

\usepackage{amsmath}
\usepackage{amssymb}
\usepackage{color}
\usepackage[utf8]{inputenc}
\usepackage{graphicx}
\usepackage{dcolumn}
\usepackage{csquotes}
\usepackage[T1]{fontenc}
\usepackage[caption=false]{subfig} 
\usepackage{cite}
\usepackage{multicol}
\usepackage{siunitx}
\usepackage{xcolor}

\begin{document}                  



\title{Simulating dark-field x-ray microscopy images with wave front propagation techniques}


\cauthor[a]{Mads}{Carlsen}{madsac@dtu.dk}{}
\author[b]{Carsten}{Detlefs}
\author[b]{Can}{Yildirim}
\author[a]{Trygve}{Ræder}
\author[a]{Hugh}{Simons}

\aff[a]{Department of Physics, Technical University of Denmark (DTU), Fysikvej, Building 311, 2800 Kgs. Lyngby. \country{Denmark}}
\aff[b]{ESRF \country{France}}

\maketitle                        

\begin{synopsis}
We show how to simulate a dark-field x-ray microscopy experiment using wave front propagation techniques and numerical integration of the Takagi-Taupin equations. We validate our approach by comparing with measurements of a near-perfect diamond crystal containing a single stacking-fault defect.
\end{synopsis}

\begin{abstract}
Dark-Field X-ray Microscopy (DFXM) is a diffraction-based synchrotron imaging techique capable of imaging defects in the bulk of extended crystalline samples. We present numerical simulations of image-formation in such a microscope using numerical integration of the dynamical Takagi-Taupin Equations (TTE) and wave front propagation. We validate our approach by comparing simulated images to experimental data from a near-perfect single crystal of diamond containing a single stacking fault defect in the illuminated volume.
\end{abstract}

\section{Introduction}
\- \\ 
Dark Field X-ray Microscopy\cite{Simons2015} (DFXM) is a full-field x-ray imaging technique similar to x-ray topography\cite{Berg1931,Lang1957}. However, unlike the latter, DFXM utilizes an x-ray objective lens placed between the sample and the detector to create a magnified image and can therefore achieve a spatial resolution better than the detector pixel size.
The spatial band width is limited by the small numerical aperture (NA ${\approx}10^{-3}$) of this objective lens. Compared to classical x-ray topography, this provides angular resolution of the scattered beam direction \cite{Poulsen2017}, which makes it possible to quantitatively measure strains and rotations of the crystal lattice by sequentially collecting images while rotating the sample and moving the objective lens. 

Traditionally, the quantitative analysis of DFXM data and the theoretical description\cite{Poulsen2017, Poulsen2021} of the method has relied on strong approximations. Most important is the \textit{kinematical approximation}, which is to omit multiple-scattering effects, and which holds for small and for highly deformed crystals. Another approximation, which is build into the geometric-optics treatment of Bragg-scattering, is that infinitesimal sub-volumes in the sample scatter according to the Bragg-law for a perfect infinite crystal, and that the intensities scattered from such sub-volumes add together incoherently. In some cases (eg.~for near-perfects crystallites and small defect-structures), however, such approximations cannot be afforded.

Here we present a method for simulating DFXM images based on wave front propagation combined with a framework that treats multiple scattering events (known as the dynamical theory of x-ray diffraction)\cite{AuthierInternationalTables}. This method is able to handle effects of coherence, dynamical scattering, aberrations of the objective lens and detector imperfections, as such it is a more realistic model of DFXM for near-perfect crystals, than the geometric optics model. This will be useful for understanding the type of contrast observed in DFXM images and can aid in experimental planning and data analysis. 
Furthermore, a number of advanced approaches to DFXM have been suggested in the literature and tested by various authors (e.g.~magnified topo-tomography \cite{Jakobsen:ks5610}, confocal Bragg-microscopy + tele-ptychography \cite{Pedersen2020}, Fourier-ptychographic DFXM\cite{Carlsen2022}). To date, the theoretical models used in these examples rely on idealized models of the underlying physics. Again, a more realistic forward model would be useful for testing the limits of their applicability. 


In this paper we discuss the steps that are required for such a simulation and present an implementation and some initial results. We compare simulated and experimental data for a near-perfect single crystal diamond containing only one single stacking fault in the imaged field of view (FOV).
	
\section{Method}
\- \\
Recent developments in DFXM are moving towards studying the dynamics of isolated defects, such as dislocations, domain walls, and acoustic waves in near-perfect single crystals\cite{Dresselhaus-Marais2021,Holstad2021}. Here, dynamical effects cannot be ignored. In these cases, the analysis of DFXM data has often relied on the \textit{weak beam approximation}, where it is assumed that even highly perfect crystals scatter approximately kinematically when the sample is rotated to the tails of the rocking curve. At this position, the bulk of the crystal does not scatter strongly. Instead, only small strained volumes near defects and surfaces contribute to the scattered intensity.
 

Much has been published on the subject of simulating x-ray topography images \cite{Taupin1964, Authier1968, Epelboin1985}. These methods are applicable in our case, but a few extra precautions must be taken due to our need for high quantitative accuracy, and the added complication caused by the objective lens. There exist useful solutions \cite{Chubar2013, Pedersen2018, Celestre2020} for simulating synchrotron sources and x-ray optical components of the kinds applied in a DFXM instrument with coherent wave-front methods. A full simulation based on coherent wave fronts is therefore within reach by combining these established methods. 



In dark-field electron microscopy (DFTEM), simulating images with dynamical diffraction effects is done routinely. Although superficially similar to DFXM, the methods cannot be transferred from one technique to the other. The most common methods in DFTEM are either Bloch wave methods that rely on the \textit{column approximation} which cannot be applied to DFXM where the scattering angles often exceed 10 degrees, or multi-slice methods that require the atomic structure of the sample to be well sampled, which is not feasible for DFXM where sample sizes often exceed 100\,\si{\micro m}.\cite{KirklandBook} 

X-ray topography methods, on the other hand, rely on the \textit{two beam approximation}\cite{Taupin1964}, which cannot be applied to electron microscopy where potentially hundreds of reflections contribute significantly to the images. 

Here we apply the formalism of the Takagi-Taupin equations, which make use of the \textit{two beam approximation} to simplify the scattering problem. This allows us to treat Bragg-scattering in an averaged way; removing the requirement to over-sample the unit cell. Instead, the sampling is limited by the divergence of the incident x-rays and the size of features in the sample. Propagation from the sample to the detector is handled by established paraxial Fourier optics methods.

We split the simulation flow into 5 steps as follows:

\begin{enumerate}
	\item \textbf{Beam}: Calculate the amplitudes of the modulated waves of the beam incident on the sample. 
	\item \textbf{Sample}: Calculate the complex scattering function throughout the sample crystal.  
	\item \textbf{Integration}: Numerically integrate the Takagi-Taupin equations to get the complex amplitudes of the scattered beam on the exit-surface of the sample. 
	\item \textbf{Propagation}: Propagate the scattered complex amplitudes through the imaging optics to the detector. 
	\item \textbf{Detector}: Interpolate the propagated field to the detector pixels and account for detector characteristics.
\end{enumerate}

\begin{figure}
	\centering
	\includegraphics[width = 0.75\columnwidth]{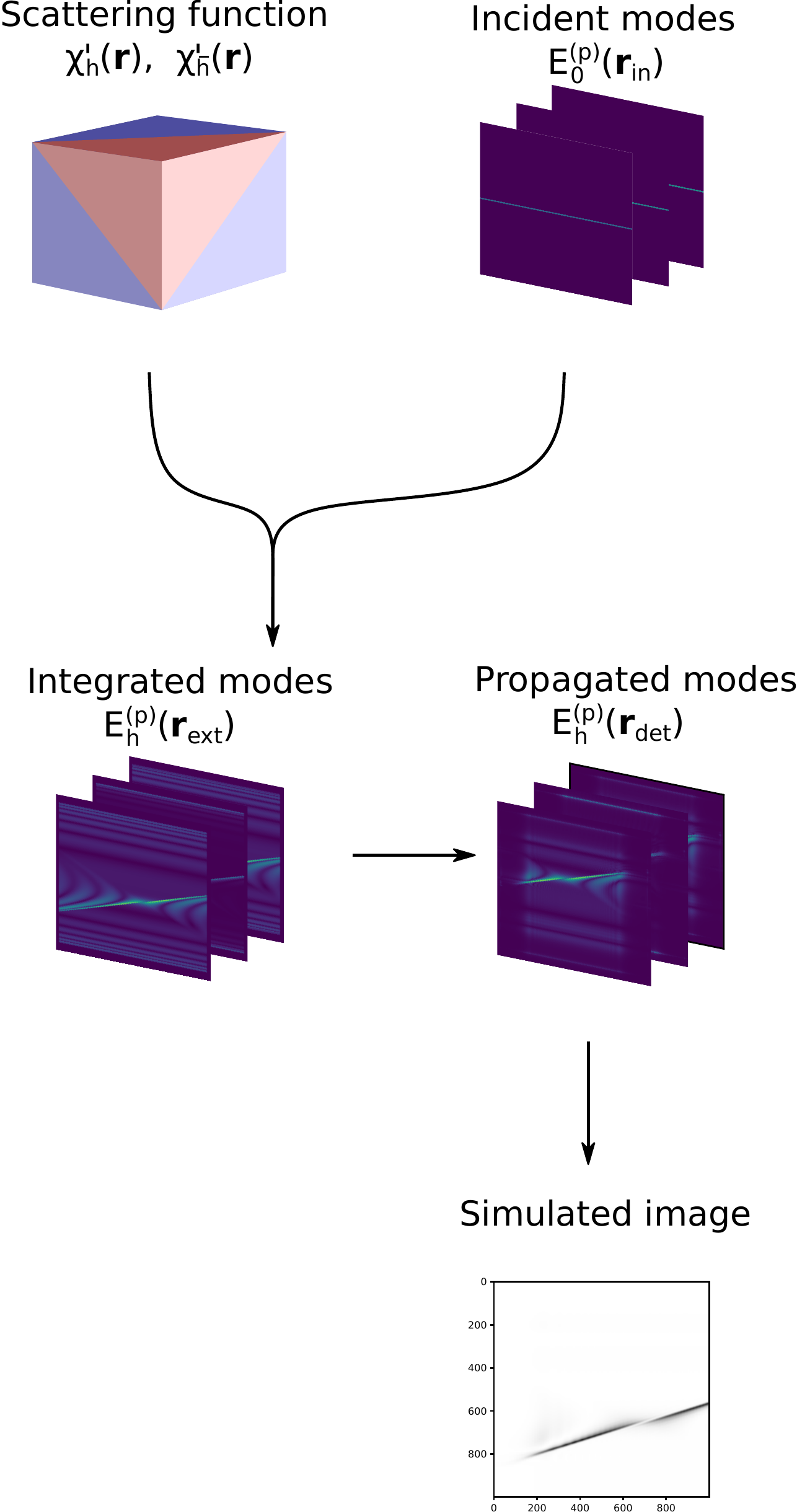}
	\caption{Schematic drawing illustrating the different steps of the simulation flow. }\label{fig:flow}
\end{figure}

A schematic drawing of the steps and flow of the simulation is shown in Fig.~\ref{fig:flow}

\subsection{Defining the beam}
\- \\ 
We work in a paraxial wave-optics formalism and use a coherent mode decomposition to describe the state of the x-ray beam at a given plane. This is given by a number of \textit{modes}, labeled with $p$: $E^{(p)}(x,y;z)$, where $E$ is the amplitude function of a modulated plane wave: $\mathbf{E} = \Re(E(\mathbf{r})\,e^{i\mathbf{k}_0\cdot\mathbf{r}- i\omega t})\,\hat{\mathbf{p}}$, where $\mathbf{r}=(x,y,z)$ is the spatial position, $\mathbf{k}_0$ is the modes' wave vector, $\hbar \omega$ is the photon energy, and $\hat{\mathbf{p}}$ is the polarization vector.

If a mode is known on a plane $z = 0$, then the modes' amplitude on another plane can be found by applying a (linear) coherent wave front propagator: 

\begin{equation}
E^{(p)}(x,y; z) = \mathcal{P}_{0 \rightarrow z}\{ E^{(p)}(x,y; 0) \}
\end{equation}

We assume that the radiation is beam-like, i.e.~that it has a limited extend in two dimensions in real space and in all dimensions in reciprocal space. The reciprocal space distribution is around a central wave vector $\mathbf{k}_{0}$, which defines the nominal direction and wavelength of the incident beam.

\subsection{Defining the sample}

In the Takagi-Taupin\cite{takagi1962a, Taupin1964, Takagi1969} approach to x-ray scattering in the two-beam approximation, the only quantities of interest are the average electric susceptibility of the crystal, $\chi_0$, and the two scattering constants $\chi_h$ and $\chi_{\overline{h}}$ that describe the cross-section for scattering and back-scattering respectively. These may be spatially modified by a displacement field $\mathbf{u}(\mathbf{r})$ of the strained crystal.
\begin{align}
	\chi'_h(\mathbf{r}) &= \exp(-i \mathbf{Q}\cdot\mathbf{u}(\mathbf{r}))\chi_h \\
	\chi'_{\overline{h}}(\mathbf{r}) &= \exp(i \mathbf{Q}\cdot\mathbf{u}(\mathbf{r}))\chi_{\overline{h}} 
\end{align}

Here $\mathbf{Q}$ is the reciprocal lattice vector of the given reflection in an undeformed reference lattice. This dependence on the displacement field explains the high sensitivity to small strains. $\chi_{h}$ and $\chi_{\overline{h}}$ can often be considered constants, but in samples with twin boundaries, there may be discontinuous jumps in the values of this function, which explains contrast observed at presumably strain-free inversion twin domain boundaries in polar materials \cite{Klapper1987}.

\begin{figure}
    \centering
    \includegraphics[width = 0.4\columnwidth]{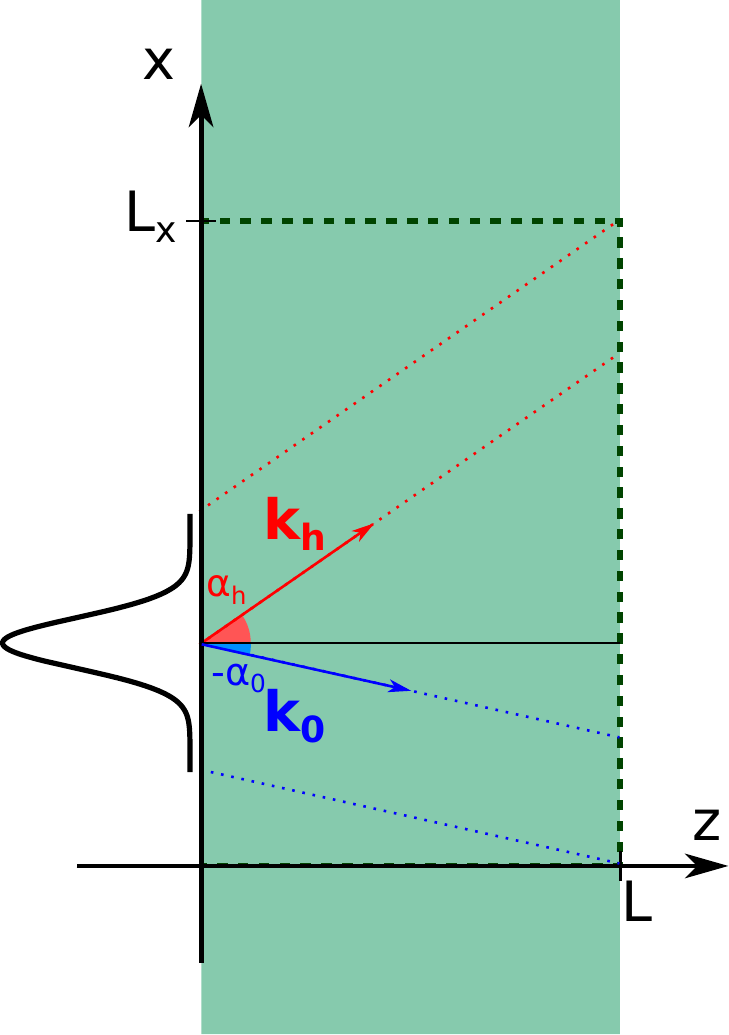}
    \caption{Scattering geometry and simulated box in the case where the scattering plane is normal to $\hat{\mathbf{y}}$. The two angles $\alpha_0$ and $\alpha_h$ fully specify the scattering geometry in this case.}
    \label{fig:vert_geom}
\end{figure}

In this paper we focus on \textit{slab-shaped} single crystals, i.e.~crystals with two parallel surfaces and infinite size in the orthogonal directions (see Fig.~\ref{fig:vert_geom}). The normal of the exit surface is called $\hat{\mathbf{n}}$.

We utilize a discrete representation of the sample structure on an orthogonal grid defined by the three directions $\hat{\mathbf{x}}$, $\hat{\mathbf{y}}$, and $\hat{\mathbf{z}} = \hat{\mathbf{n}}$ and corresponding step sizes $d_x$, $d_y$, and $d_z$. It is necessary that the surface normal of the crystal slab is parallel to the $\hat{\mathbf{z}}$-axis, but we make no requirement on the last free rotation. The number of grid points in each direction will be labeled $N_x$, $N_y$, and $N_z$, giving a total size of the simulated domain of $L_x = d_x(N_x-1)$, $L_y = d_y(N_y-1)$, $L = d_z(N_z-1)$. $L$ is the thickness of the simulated crystal slab.

The complex value of these scattering functions needs to be known with high resolution. For simple test-cases, where the displacement field is given by an analytical expression, this is not a point of concern. If, however, the displacement field is generated by a numerical simulation, then it needs to be computed with sufficient resolution to at least match the resolution of the experiment (${\approx}50\,\si{nm}$) throughout the volume of the sample. The size of the sample can be up to several hundred micrometers. 

For highly deformed samples, the scattering function contains a phase factor of the shape $e^{i\mathbf{Q} \cdot \mathbf{u}}$. In order to limit the phase variation between adjacent voxels to less than $2\pi$, step-sizes must be bellow $|\mathbf{\nabla} \mathbf{u}|/(2\pi|Q|)$, which means small steps must be used for highly deformed samples. In samples with nanometer-sized domains or other small stuctures, all structural features must be resolved.


\subsection{Integrating the Takagi-Taupin equations}

Scattering of X-rays by a deformed crystal is treated in the formalism of the symmetric Takagi-Taupin equations (TTEs) \cite{Vartanyants2001}, which for this particular choice of parameters takes the shape:

\begin{equation}
\begin{split}
2\left(\mathbf{k_{0}}\cdot\mathbf{\nabla}\right) E_0^{(p)} &= -ik{}^2\chi_0 E_0^{(p)}
 -ik{}^2C\chi_{\overline{h}}'E_h^{'(p)}   \\ 
2\left(\mathbf{k_{h}}\cdot\mathbf{\nabla}\right) E_h^{(p)} &= -ik{}^2(\chi_0 + \beta)E_h^{(p)} -ik{}^2C\chi_{h}'E_0^{(p)}  
\end{split}\label{eq:TT} 
\end{equation}

where $\beta = 2\sin(2\theta)\phi$ is a measure of the misorientation away from the \textit{vacuum} Bragg-condition, $\phi$ is the rocking angle and $C$ is the polarization factor. $E_0^{(p)}$ are the modes of the plane wave decomposition of the incident beam and $E_0^{(p)}$ are the corresponding modes of the scattered beam, which is given relative to the wave vector $\mathbf{k}_h = \mathbf{k}_0 + \mathbf{Q}$. 

The geometry of the problem is set by the shape of the incident beam, the vectors $\mathbf{k}_0$ and $\mathbf{Q}$, as well as the choice of a computational grid. The plane spanned by the two vectors $\mathbf{k}_0$ and $\mathbf{Q}$ is called \textit{the scattering plane}.

When the scattering plane is normal to the $\hat{\mathbf{y}}$-axis, the TTEs decouple into a set of 2D problems that can be solved slice-by-slice. If we are free to choose the orientation of the computational grid, we can always choose a geometry where this becomes the case.

When $\mathbf{Q}$ is parallel to the surface of the crystal, we say that we have a \textit{symmetric Laue geometry}. In this case we can choose an orientation of the  computational grid where $\mathbf{Q}||\hat{\mathbf{x}}$, where the TTEs take a particularly simple form.



In order to solve the TTEs, we impose zero Dirichlet boundary conditions in the two transverse dimensions, $x$ and $y$. These boundary conditions require the sample grid to be large enough to fit the entire Borrmann triangle extending from every point where the incident beam amplitude is non-zero. This is fulfilled if the non-zero part of the amplitude is fully contained in the rectangle defined by: (see Fig. \ref{fig:vert_geom})

\begin{equation}
\begin{split}
\max(0, &Lk_{0x}/k, Lk_{hx}/k)< x <\\ 
& \min(L_x, L_x+Lk_{0x}/k, L_x +Lk_{hx}/k) \\
\max(0, &Lk_{0y}/k, Lk_{hy}/k)< y <\\ 
& \min(L_y, L_y+Lk_{0y}/k, L_y+Lk_{hy}/k), 
\end{split}
\label{eq:geom_constr}
\end{equation}
where $L_x, L_y$ are the lengths of the simulated domain in the $x$ and $y$ directions and $L$ is the thickness of the crystal. For a given incident beam and $\mathbf{Q}$, this sets a minimum on the size of the simulated domain in the two transverse directions. 

With the sample structure and the boundary conditions given, the TTEs constitute an initial value problem, where the initial value is the amplitude of the incident x-rays on the $z=0$ surface. This can be solved by an appropriate finite difference scheme to yield the amplitudes of both the transmitted and scattered beams on the exit-surface $z = L$. Details of the applied finite difference scheme are given in \cite{Carlsen2021}. \\

\subsection{Propagating through the optics }




In DFXM, the scattered beam at the exit surface of the sample is imaged onto a detector using an objective lens in a magnifying geometry. The propagation through the lens is challenging to simulate due to the inherent thick-lens behavior of the Compound Refractive Lens (CRL) which is typically used as the objective lens. For this, we use a computational approach where each lens in the CRL is treated as a thin lens and a paraxial FFT-propagator is used to propagate the wave front between each lens. 

To do this, we make use of a method for  propagating wave fronts with rapidly varying quadratic phases that come from the transmission function of the lenses that treats the quadratic component analytically.\cite{Ozaktas1996, Chubar2019} These methods have previously been used to model thick CRLs like the one used in this study \cite{Pedersen2018,Celestre2020}.

It is useful to introduce a new optical axis ($\hat{\mathbf{z}}_i$) aligned with the average wave-vector of the scattered beam, $\mathbf{k}_h$. To bridge the gap between these two coordinate systems, we project the values from the exit surface of the crystal to the $z = 0$ plane of the new coordinate system. 

The inherent near-field nature of the imaging geometry, which is determined by the fact that the FOV is as large as the aperture of the objective lens, is handled by first multiplying the field with the near field phase factors: 

\begin{equation}
	\phi_{\mathrm{nf}} = \exp\left[ \frac{i\pi(x_i^2 +y_i^2)}{\lambda d_1} \right]
\end{equation}


The need to over-sample this function can effectively limit the size of the FOV that is possible to simulate for a given pixel-size.

\subsection{Detector characteristics}

With the mode amplitudes on the detector-plane given, we now need to interpolate these values to the detector pixels and incoherently sum over the modes of the incident beam. 

If the purpose of the simulation is to estimate the resolution or to create a data set to be used to test the data analysis procedures, one should remember to include non-ideal behavior introduced at the detector. The most important effects are the incoherent point spread of the detector, background signal, counting noise and non-linear response.

The detector used here is an indirect detector composed of a scintillator crystal ($25\,\si{\micro m}$ thick gadilinium gallium garnet), optical microscope (Mitutoyu M Plan Apo $10\times$, NA=0.28 and tube lens), and pco.edge 5.5 sCMOS camera with pixel size $6.5\,\si{\micro m}$. Contributions to the incoherent point spread function arise from the diffraction limit due to the finite NA of the optical microscope, the finite thickness of the scintillator (especially when it is larger than the depth-of-focus of the optical microscope), scattering within the scintillator and the optical microscope, and aberrations.

Counting statistics/readout noise can be the critical factor when imaging small grains or weak reflections. Non-linear response (saturation) might be quite important for perfect crystals, as the images have interesting features over a very large dynamic range.

\section{Comparison with experiment}
\- \\ 
To test our approach, we simulate a section-topography type experiment with a near-perfect single-crystal diamond slab of thickness $300\,\si{\micro m}$  containing a single stacking fault defect. The sample is imaged in a symmetric Laue geometry in  a $\{111\}$-reflection with [110] entrance and exit surfaces. 

The investigated defect is a stacking fault, which arises by the addition or removal of a single close-packed plane of atoms in the FCC parent-lattice of the diamond. The fault vector is of the family  $\mathbf{b}_{\mathrm{sf}} =\frac{1}{3}\langle 111\rangle$ which is not a translational symmetry of the fcc lattice. These planar defects are bounded by the surfaces of the crystal and Frank-type partial dislocations\cite{Frank1951,Kowalski1989}. In the described experiment, the edges of this defect lie outside the Borrmann triangle and the effects of the strain originating at the edges can be ignored. This allows us to treat the stacking fault as an infinite planar defect (on one side $\mathbf{u}(r) = 0$, on the other $\mathbf{u}(r) = \mathbf{b}_{\mathrm{sf}}$). In the Takagi-Taupin description, the stacking fault thus becomes a discrete jump in the phase of the scattering function of magnitude $2\pi[hk\ell] \cdot \mathbf{b}_{\mathrm{sf}} = \pm 2/3 \pi$ when imaged using a [111] reflection not orthogonal to the stacking fault normal. \cite{Klapper1987}

This constitutes a good test sample, as the defect (aside from the 12 possible orientations of the defect) only has a single continuous degree of freedom: the position of the stacking fault along the normal. Furthermore, diamond is one of a few materials where macroscopic crystals with very low defect density are available. Provided our method captures all the relevant physics, we should therefore be able to perfectly recreate the experimental data.

The dynamical scattering patterns produced by isolated stacking faults in diamonds have previously been studied in detail by classical x-ray topography.\cite{Kowalski1989}

Experiments were carried out at the ESRF dark-field x-ray microscopy beamline, ID06-HXM \cite{Kutsal2019}. A Si(111) Bragg-Bragg double crystal monochromator selected x-rays with photon energy 17\,\si{keV} from the undulator source. The spot size of the beam on the condenser lens is limited by a slit in the vertical direction to 0.2\,\si{mm}. The small effective source size of the insertion device and long distance from the source to the condenser lens (~50\,\si{m}) means we can assume that the incident x-rays are fully coherent in the $x$ direction. The objective lens consists of 70 individual biconcave Be lenses of apex curvature 50\,\si{\micro m}. The CRL is further modified with a 100\,\si{\micro m} square aperture after the last lens

\begin{figure}
    \centering
    \includegraphics[width = \columnwidth]{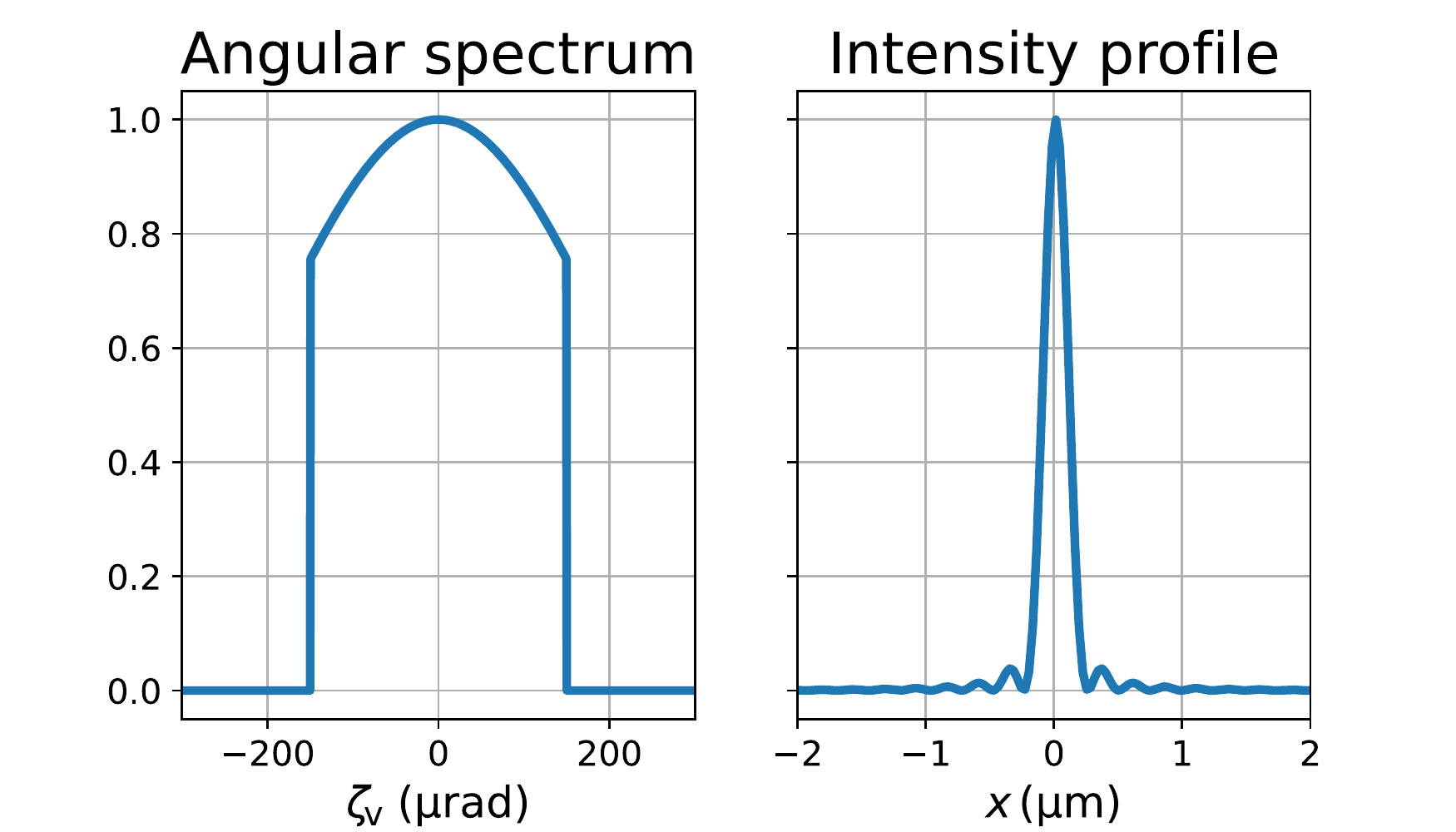}
    \caption{Vertical profile in a) reciprocal and real b) space of the beam used in the simulation of this paper. }
    \label{fig:beam}
\end{figure}

To describe the incident x-rays, we use only one coherent mode for each wavelength and 41 different wavelengths covering a relative energy spread of $6\cdot 10^{-4}$ in total. For each energy component, the amplitude function is taken to be the Fourier transform of an aberration-free 1D condenser lens with a Gaussian-shaped absorbing aperture and a hard cut-off at $140\,\si{\micro rad}$. Consequently, the angular spectrum in the transverse direction is a top-hat profile with a small Gaussian smoothing at the edges, whereas the spatial profile is a diffraction-limited focal spot with some ringing due to the hard cut-off (see Fig.~\ref{fig:beam}). 

\begin{figure}
    \centering
    \includegraphics[width = \columnwidth]{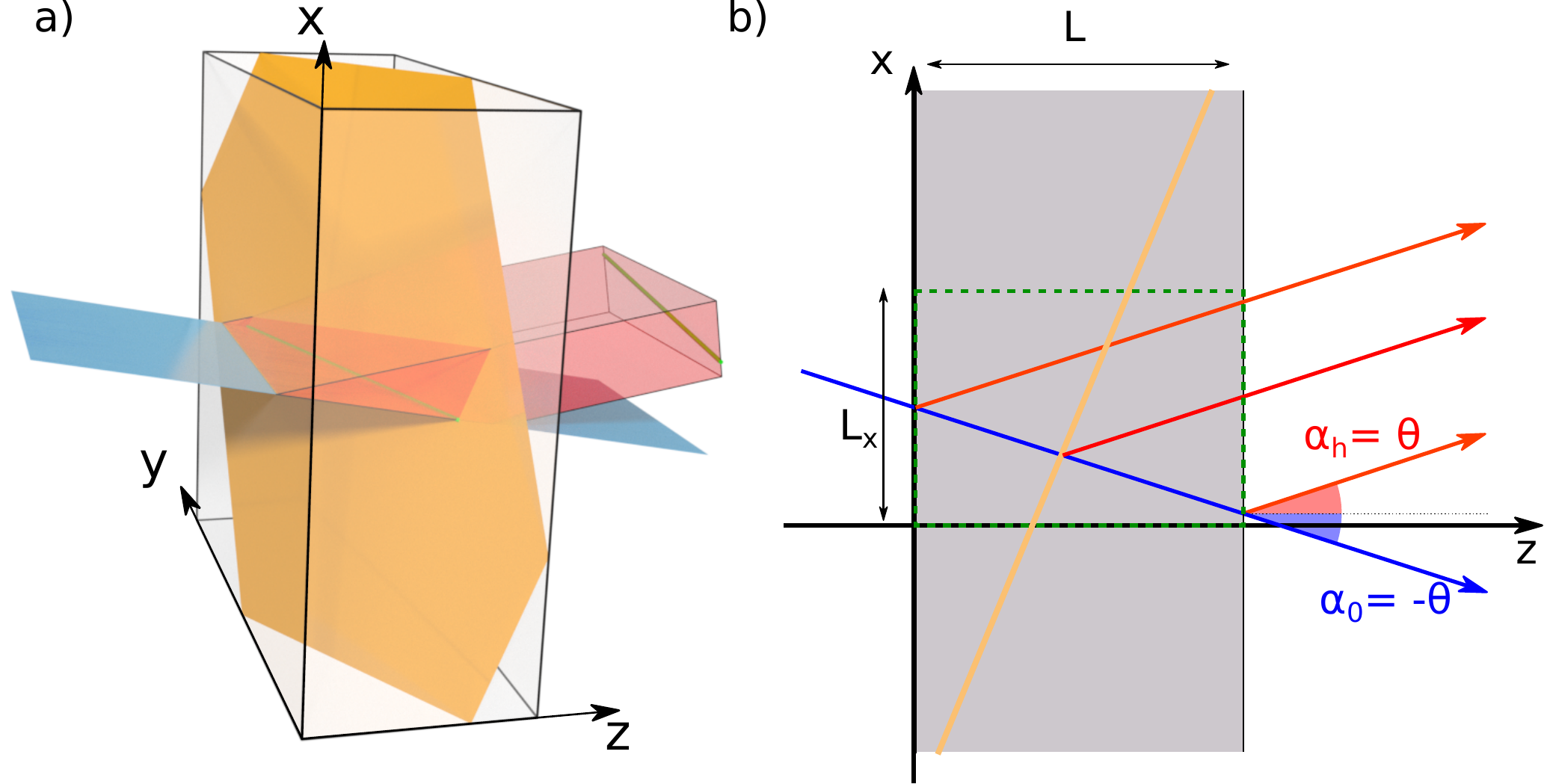}
    \caption{a) 3D sketch of the simulated geometry where the blue sheet represents the incident/transmitted x-ray beam, the red represents the scattered rays and the yellow plane represents the stacking fault. b) 2D slice of the simulated geometry showing the same features as a) with definitions of useful angles and distances.}
    \label{fig:geom_sketch}
\end{figure}

In the experimental realization we make the observations in a \textit{laboratory frame} defined by $\hat{\mathbf{x}}_{\mathrm{lab}}$, $\hat{\mathbf{y}}_{\mathrm{lab}}$, and $\hat{\mathbf{z}}_{\mathrm{lab}}$ where $\hat{\mathbf{z}}$ is parallel to $\mathbf{k}_0$. In this experiment the scattering is \textit{vertical} in the lab frame, meaning that $\hat{\mathbf{y}}_{\mathrm{lab}}$ is normal to the scattering plane. We choose the integration grid such that $\hat{\mathbf{y}} = \hat{\mathbf{y}}_{\mathrm{lab}}$.

The simulation used a grid of $2048\times2048\times3001$ points with corresponding step sizes of 60\,\si{nm}, 60\,\si{nm}, and 100\,\si{nm} respectively. This matches the 300\,\si{\micro m} thickness of the sample and the ${\approx}100\,\si{\micro m}$ FOV in the $y$-direction. The large size in the x-direction was necessary to satisfy the constraints of Eq.~\eqref{eq:geom_constr}. Step sizes in $x$- and $y$-directions are chosen to be larger that the resolution of the final images, and the step size in $z$ is chosen such that the integration error of the finite-difference scheme is lower than the noise level of the final images. The execution time is dominated by the integration of the Takagi-Taupin equations, which took 3.5 hours per mode running on a single core. The simulation code has not been optimized for performance.

\subsection{Nearfield measurements}

The microscope used for the experiment provides the possibility to additionally carry out traditional x-ray topography by placing a detector closely behind the sample (40\,\si{mm} in the examples shown here) without using the objective lens. This is useful for alignment and for low resolution characterization of the sample.

\begin{figure}
\centering
\includegraphics[width = \columnwidth]{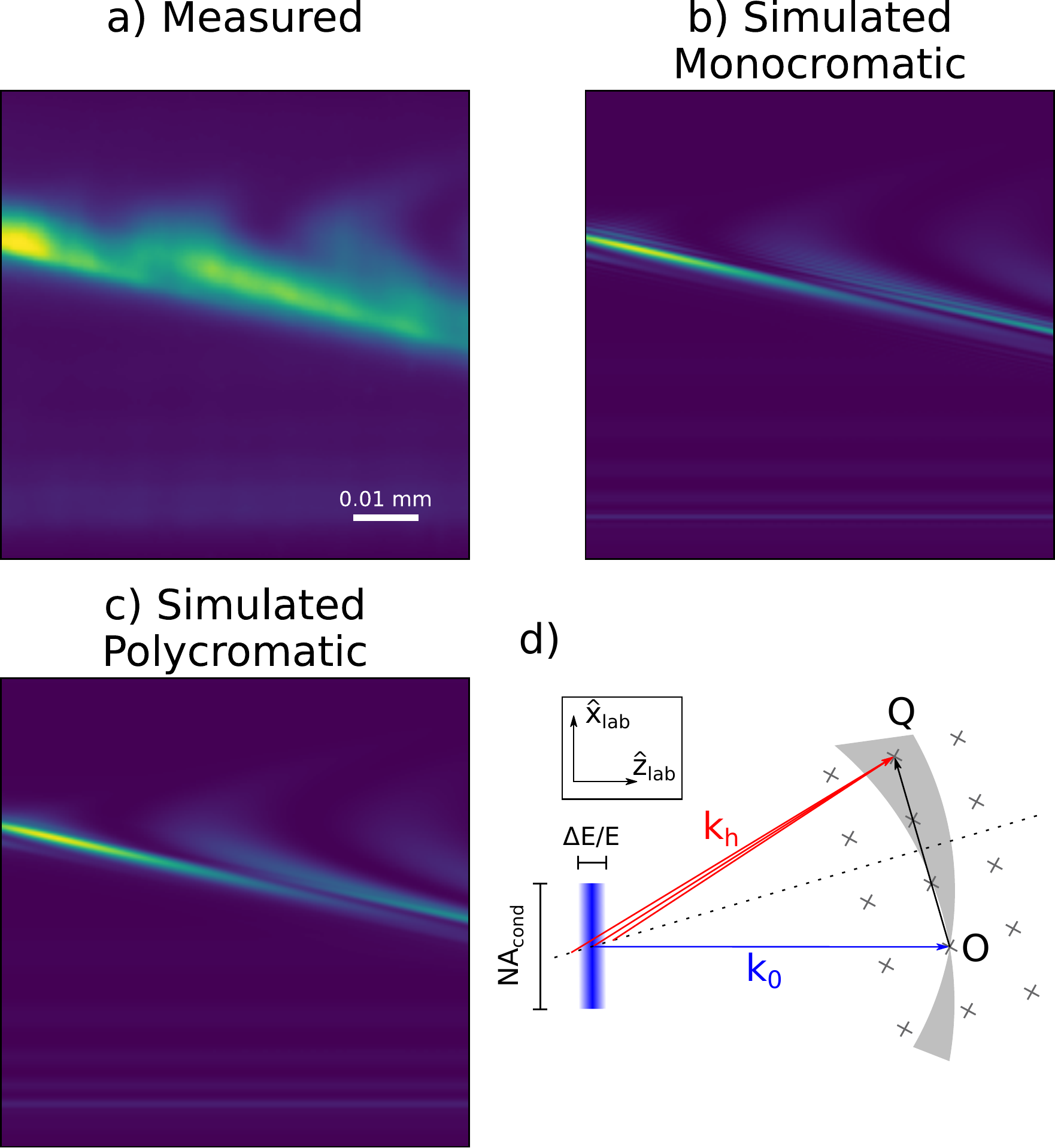}
\caption{X-ray section topography of the single diamond (110) sample containing a single stacking fault defect in the imaged volume, imaged at a propagation distance of 40\,\si{mm}. a) Experimental image, b) Simulated image with one coherent mode. c) Polychromatic simulated image. d) Sketch of the scattering geometry in reciprocal space in the laboratory frame. The incident x-rays consist of a continuum of $\mathbf{k}_0$-vectors (blue rectangle) due to the divergence and finite bandwidth of the incident beam this smears out the Ewald's sphere to a shell. For a given rotation of the crystal only points on the dotted line (the bisection of the $\mathbf{Q}$-vector) satisfy the Bragg-condition. Different energy components scatter at slightly different angles.}
\label{fig:nearfield}
\end{figure}

For a single coherent mode (Fig.~\ref{fig:nearfield} b) this propagation distance causes recognisable Fresnel-diffraction fringes around sharp features in scattered field. In a polychromatic simulation (Fig.~\ref{fig:nearfield} c), these fringes are blurred out. The difference in wavelength (which causes a small difference in the free-space propagator) is not sufficient to explain this blurring. Rather, the broadening is caused by the slight difference in scattering angle of the different energy components in the incident beam (Fig.~\ref{fig:nearfield} d). The features in the simulated image, however, are not as wide as in the measured data (Fig.~\ref{fig:nearfield} a). This is likely explained by the incoherent point spread of the detector.

\begin{figure}
\centering
\includegraphics[width = \columnwidth]{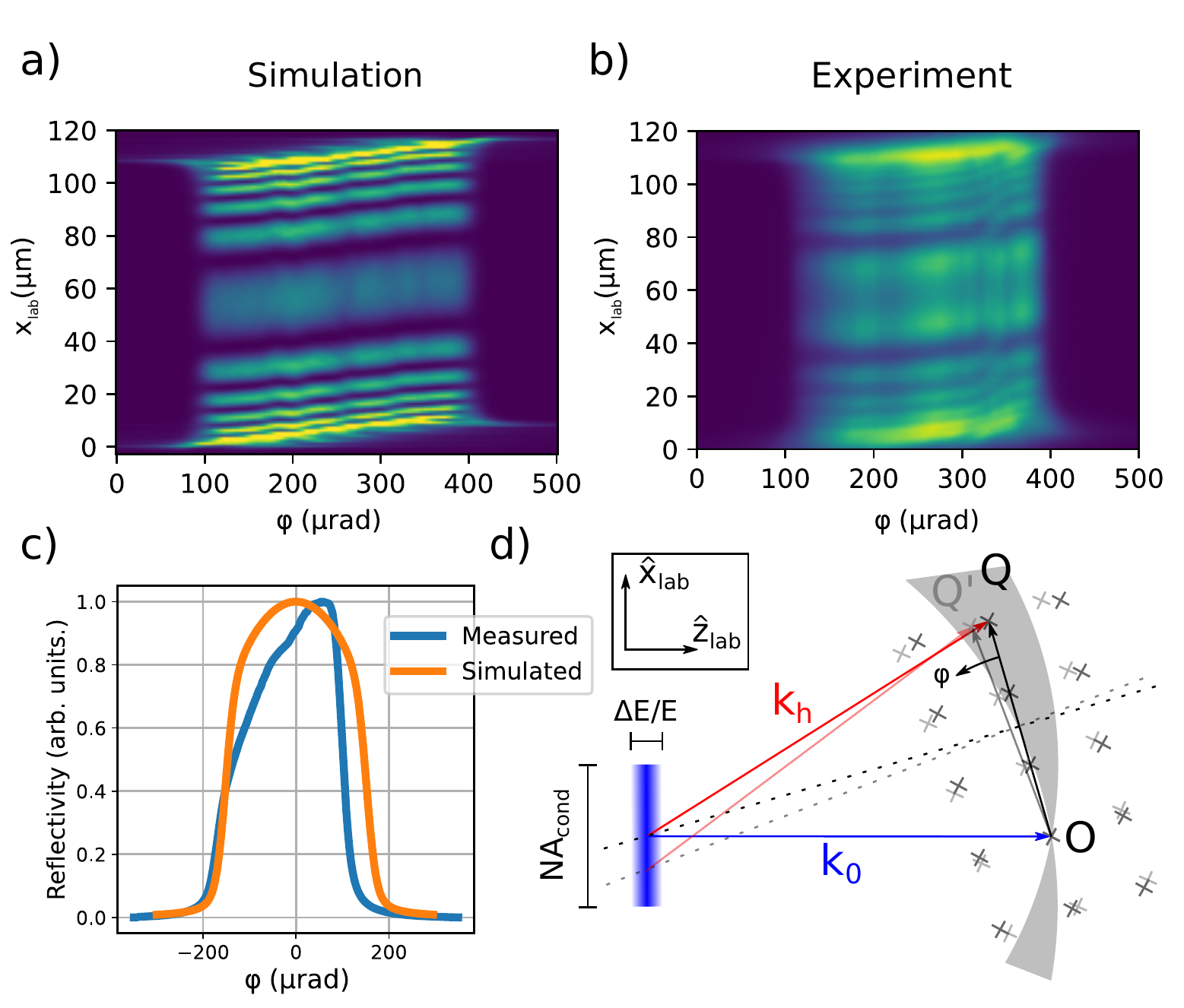}
\caption{ a) Simulated spatially resolved rocking curve of a slice of the diamond crystal away from the stacking fault using an aberrated model of the condenser lens. b) Measured spatially resolved rocking curve of a slice of the diamond crystal away from the stacking fault. c) Measured and simulated spatially integrated rocking curves of a diamond single crystal in the region containing a single stacking fault. d) Sketch of the scattering geometry in reciprocal space in the laboratory frame. The incident x-rays consist of a continuum of $\mathbf{k}_0$-vectors (blue rectangle) due to the divergence and finite bandwidth of the incident beam this smears out the Ewald's sphere to a shell. For a given rotation of the crystal only points on the dotted line (the bisection of the $\mathbf{Q}$-vector) satisfy the Bragg-condition. When the crystal is rotated by an angle $\phi$ (thereby rotating $\mathbf{Q}$ into $\mathbf{Q}'$), this line is moved and a different part of the incident spectrum satisfies the Bragg-condition. This leads to scattering in a different direction in the laboratory frame (comparing red and pink arrows).}
\label{fig:rock_image}
\end{figure}

The vertical divergence of the condensed line beam is large compared to the intrinsic width of the dynamical rocking curve of the diamond sample. The crystal therefore acts as an analyzer when rocked in the condensed line-beam (see Fig.~\ref{fig:rock_image}).
The width of the rocking curve is largely determined by the divergence of the incident beam, but the finite energy band-width blurs the sharp cutoff caused of the condenser lens' physical aperture. The asymmetry of the measured rocking curve (Fig.~\ref{fig:rock_image} c) reveals a misalignment of the condenser lens.

High-frequency defects in the condenser lens are visible in the spatially resolved rocking curve of perfect parts of the crystal (see Fig.~\ref{fig:rock_image}b ) as vertical stripes. The stripes are blurred along the rocking-angle direction due to the finite bandwidth of the incident x-rays --- an effect which is confirmed by the simulations shown in Fig.~\ref{fig:rock_image}a.

When the crystal is rocked, a different part of the spectrum of the incident beam will be in the Bragg-condition and therefore the sample will scatter in slightly different directions as a function of the rocking angle $\phi$ (sketched in Fig.~ \ref{fig:rock_image}d ). Due to the finite propagation distance from the sample to the detector, this change in angle translates to a change in position of the measured image. This mix of position- and angular information, which is avoided by the use of an objective lens, is unavoidable in x-ray topography methods due to the finite propagation length, but it is exaggerated in this study due to the relatively large propagation distance and large vertical divergence compared to more usual topography experiments.

\subsection{DFXM images}

To simulate the DFXM images, we use a model of an ideal CRL. The physical aperture of the CRL is defined by a  $0.1 \times 0.1 \,\si{mm^2}$ square absorbing slit placed at the exit of the lens. The only fitted parameters for the whole simulation are the relative positions of the sample, lens, and detector as well as the noise level of the detector. The experimental images are overexposed (saturated) at the direct image of the stacking fault, therefore we here choose a color map that clips the highest intensities in the simulated images.

\begin{figure}
\centering
\includegraphics[width = \columnwidth]{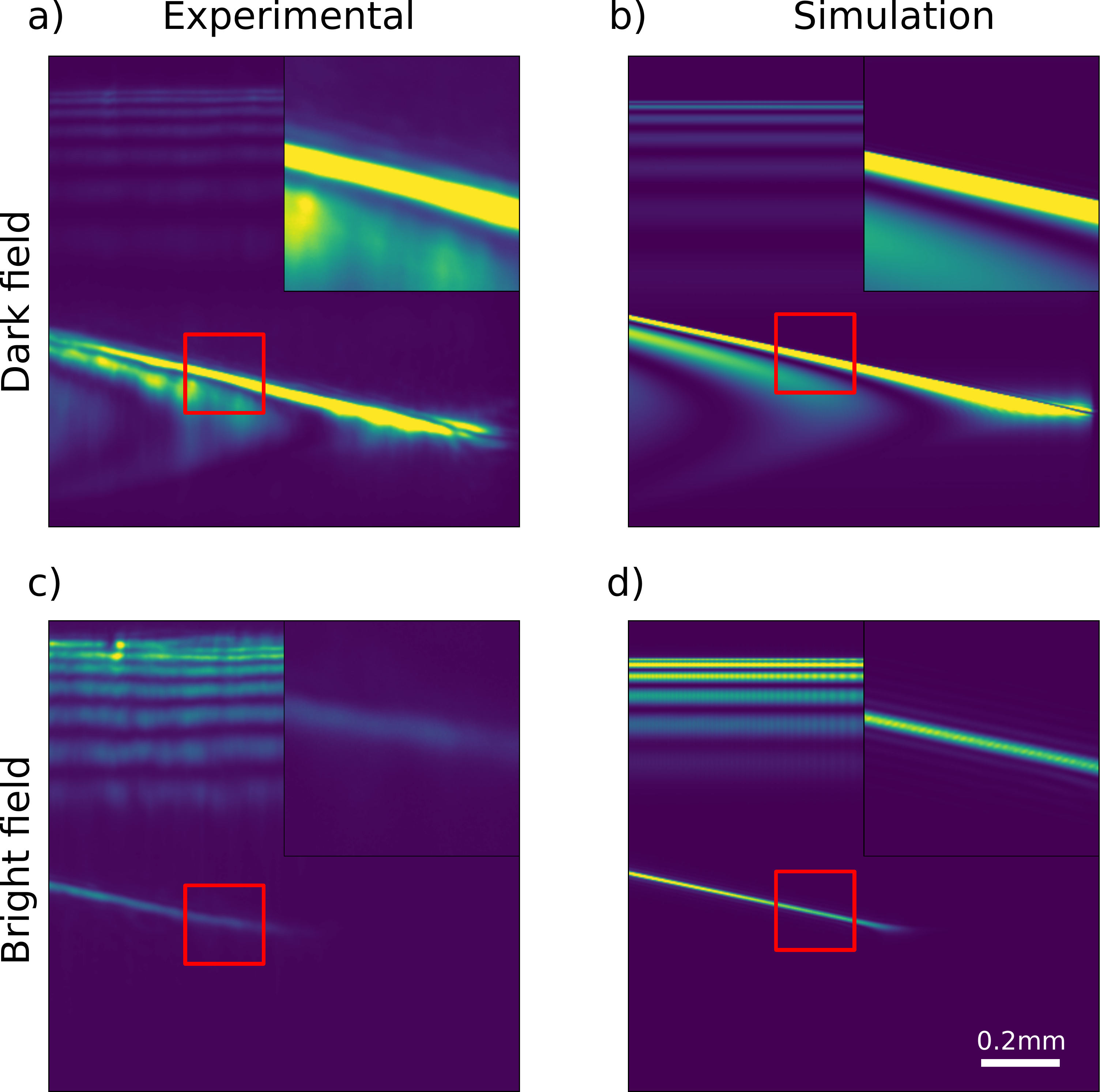}
\caption{Measured (a,c) and simulated (b,d) DFXM images of a stacking fault defect in a diamond single crystal. (a,b) show images where the objective lens is placed in the center of the diffracted beam. (c,d) show images where the lens is displaced from the diffracted beam leaving the bottom half of the FOV in the dark field. }
\label{fig:images}
\end{figure}	


As can be seen in Fig.~\ref{fig:images}, the DFXM simulation qualitatively recreates the features of the experimental images. However, there are a number of deviations: 

\begin{itemize}
\item
We underestimate the magnification of the imaging set-up by about 5\%, which leads to an incorrect scaling of the images. This is most likely due to a small deviation of the apex radius of curvature from the nominal value of 50\,\si{\micro m} in the individual lenses of the CRL used as objective lens.

\item
The simulated images contain a smaller number of Borr\-mann fringes (the horizontal stripe features) than the measured data.
We attribute this to the known high sensitivity of the spacing of Borrmann fringes to small macroscopic strains-gradients.\cite{Rodriguez-Fernandez2021} 

\item
The simulated images have a regular pattern of vertical streaks close to the right hand side of the images. These are due to Fresnel diffraction from the hard edge of the square aperture in the objective lens. This is likely an artifact of the assumption of perfect transverse coherence in the horizontal direction or of the perfectly sharp edges of the aperture that are somewhat jagged in practice. 

\item
The measured images contain noise with the appearance of vertical streaks and speckle-like features close to the brightest features,. This can be explained either by the aberrations in the condenser lens or in the objective lens, as will be discussed later.

\end{itemize}

In Fig.~\ref{fig:images} c,d) we simulate an image where the objective lens is displaced from the center of the scattered beam such that rays that are specularly reflected fall outside of the aperture of the objective lens in the bottom part of the displayed ROI. In that region, only diffusely scattered x-rays will contribute to the image. This results in the disappearance of the dynamical features, while the direct image of the stacking fault can still be seen. In visible light microscopy, this is referred to as ``dark-field contrast''.

In the geometric optics treatment, weak beam contrast is explained by the presence of small regions where the lattice is strained and rotated away from the average lattice. In these regions, rays are scattered if they satisfy the exact Bragg condition for the deformed lattice. A stacking fault is in principle a perfectly sharp defect with no spatial extent so no such region exists. The appearance of weak beam contrast therefore illustrates the inability of the geometric model to handle diffraction effects that are important when describing scattering from small structures with a characteristic size on the order of the wavelength or smaller.


Since the stacking fault is thought to be a perfectly sharp defect, the width of the image of this defect can serve as a rough estimate of the resolution of the instrument. The stacking fault is a 2D feature, and therefore the width of the image is not only determined by the resolution of the imaging optics, but also by the projection of the part of the stacking fault illuminated by the sheet beam along the scattered beam direction.

\begin{figure}
	\centering
	\includegraphics[width = 1\columnwidth]{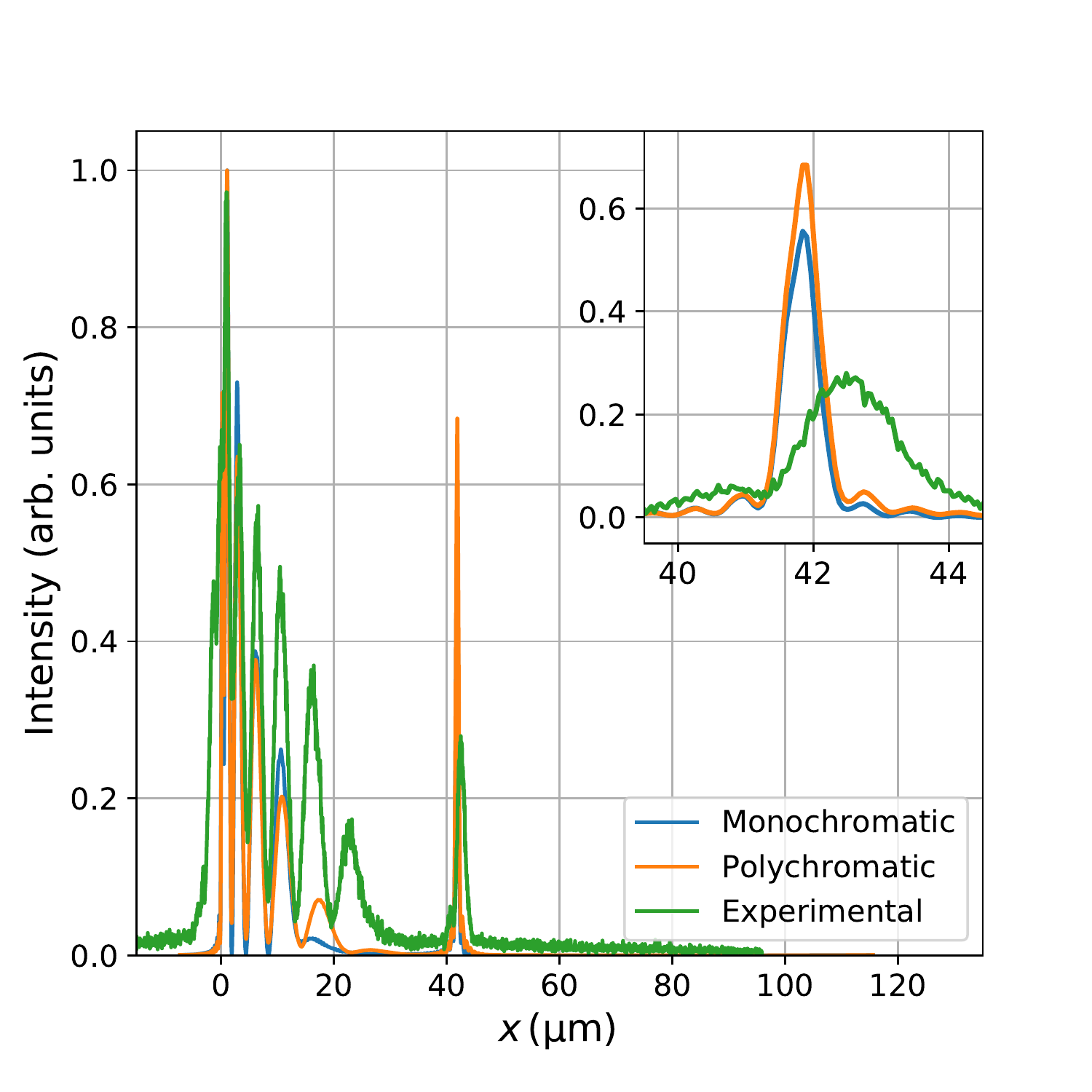}
	\caption{Comparison of the width of the dark-field image of the stacking fault as seen in Fig.~\ref{fig:images} c,d). The $x$-axis refers to distances in the sample plane coordinates.}
	\label{fig:stack_fault_DF_image_width}
\end{figure}

In Fig. \ref{fig:stack_fault_DF_image_width} (insert) we compare the width of this feature in the experimental images with the simulated images. We see that the polychromaticity does not contribute significantly to the width of the feature in the simulations. Previous studies of the chromatic aberrations in CRLs, using the same computational approach as we apply here,  also find that the chromatic aberration only add a small part to the point spread of the imaging optics.\cite{Pedersen2018}

The experimental image is about 0.5\,\si{\micro m} wider in the demagnified sample-plane coordinates than that predicted by the simulations. We believe that the resolution of the experiment is degraded by aberrations in the CRL lenses. 

\subsection{Aberrated lenses}

So far, we have ignored the effect of the aberrations in the lenses.
In transmission images\cite{Lyatun2020} and Bragg-images taken without the condenser lens, short-wavelength aberrations are known to cause strong speckle-like noise in the final images. The apparent absence of this noise in DFXM images is surprising at first. However, as previously observed, this noise is averaged out when the imaged field is only partially coherent\cite{Falch2019,Carlsen2022}. Normally we think of the dynamically scattered x-rays as highly coherent as the Bragg scattering effectively collimates the incident radiation, but this argument does not consider the polychromaticity of the incident radiation. 

While CRLs have been shown to be nearly achromatic over the bandwidth of the monochromatized beam \cite{Pedersen2018}, Bragg scattering is not: A higher energy component of the incident beam scatters at a smaller angle and vice versa, as sketched in Fig. \ref{fig:nearfield}d. Since the incident beam has a large vertical divergence (compared to both the energy bandwidth and the Darwin width) set by the aperture of the condenser lens, the integration over energies corresponds to integrating over a small spread of angles of the scattered beam. This integration averages out the high-frequency parts of the aberrations in the vertical direction, leaving features elongated in the vertical direction. A relative energy spread of $1.0\cdot 10^{-4}$ corresponds to an angular difference of $1.0\cdot 10^{-4}\,\si{rad} \cdot \tan\theta $ which gives 8\,\si{\micro m} at the lens-plane --- comparable to the average grain size (15\,\si{\micro m}) of the O30H grade Beryllium \cite{Lyatun2020} used in our CRL. 

\begin{figure}
	\centering
	\includegraphics[width = \columnwidth]{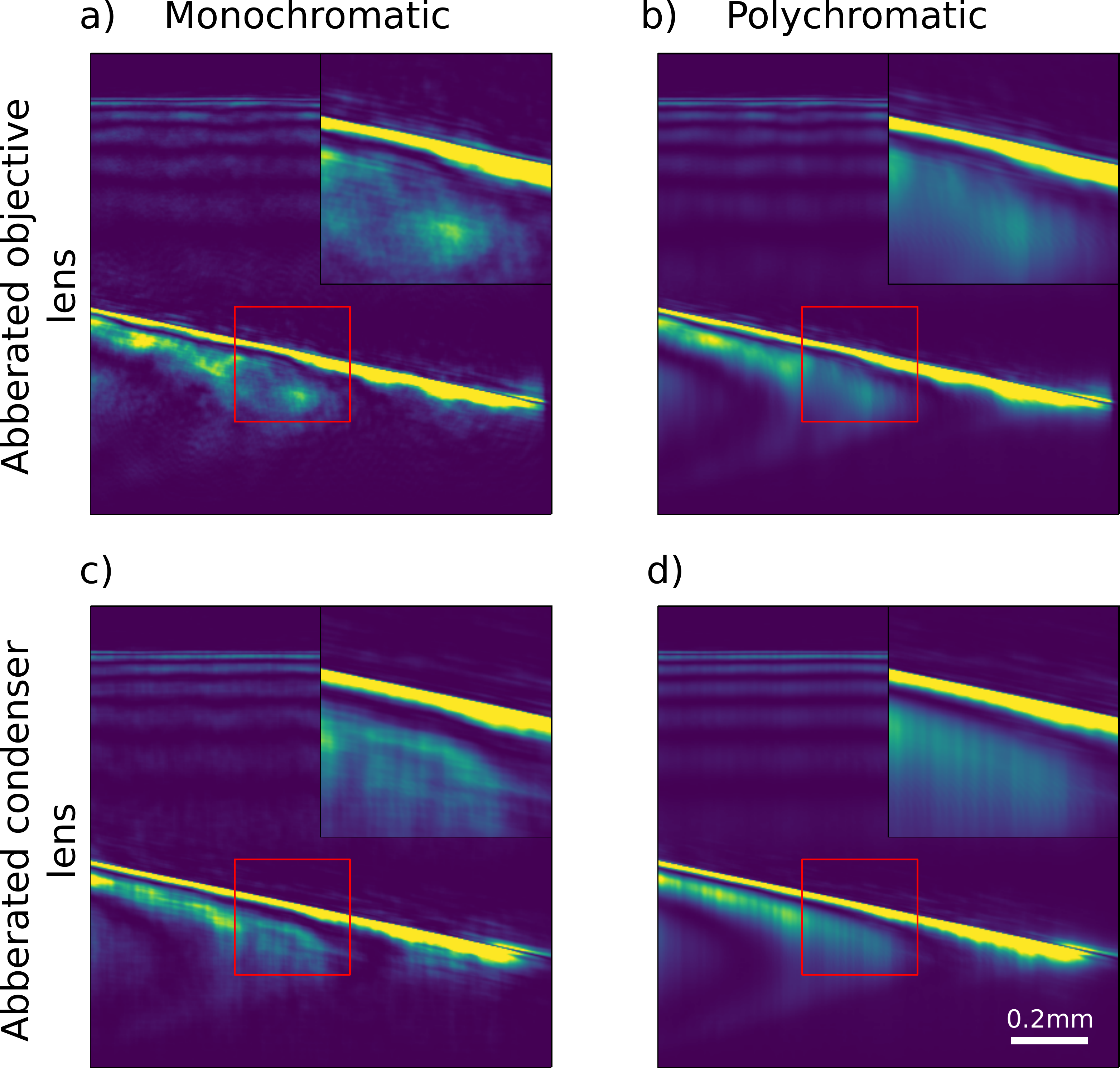}
	\caption{a) Monochromatic and b) polychromatic simulation using an aberrated objective lens. c) Monochromatic and d) polychromatic simulation using an aberrated condenser lens. The scale bar refers to distances on the detector. }
	\label{fig:aberrated_lens_simulation}
\end{figure}

This effect is demonstrated in Fig.~\ref{fig:aberrated_lens_simulation} a,b, where an aberrated lens is constructed by multiplying the wave front by an aberration function at the position of the first lens and the last lens in the CRL. The aberration functions used here are pure phase objects and are made by randomly placing a number of circles of random size. The amplitude and number of circles is chosen to make the simulated and measured images similar. The partially averaged speckle noise has the appearance of vertical stripes and is qualitatively similar to that observed in the real images (Fig. \ref{fig:images} a). 
In a typical experiment, 
The authors kindly acknowledge the use of beamtime from the European Synchrotorn. M.A.we do not acquire sufficient data to uniquely determine the aberrations, but an effective aberration function can be recovered using Fourier ptychography \cite{Carlsen2022}. The qualitative similarity between simulated and measured images confirms that the vertical stripe artifacts seen in DFXM images of highly perfect crystals can be explained by high-frequency errors in the objective lens, which we know to be present.

In Fig. ~\ref{fig:aberrated_lens_simulation} c,d) we investigate the effect of similar aberrations, to those used in the objective lens, in the condenser lens. Once again, averaging over the energy bandwidth significantly reduces the strength of the noise and results in vertical stripes. In this case the stripes are unbroken and can be followed from the top to the bottom of the image, in contrast to the noise observed in the real images and with an aberrated objective lens.

\section{Conclusion}
\- \\ 
DFXM is based on well-known physics and we can predict the images it will produce --- if we know the structure of the sample. It is possible to simulate the full FOV of the prototypical DFXM instrument at ID06-HXM at the ESRF.  

Comparing our simulations and experimental findings from a near-perfect single crystal diamond suggest that most deviations between our simulations and the observed images can be explained by non-ideal behaviour of the lenses. This suggest that the performance of DFXM instruments is critically limited by the quality of the objective lens. 

In general, we do not have a sufficiently accurate model of the sample structure to do full simulations of the DFXM experiment, and the data collected in a typical experiment is not enough to build a full 3D model of the sample at sufficiently high resolution. Nevertheless, simulations like the ones shown here should prove useful for evaluating possible upgrades of the instrumentation and to qualitatively study the type of contrast observed from different types of defects in near-perfect single crystals, such as isolated dislocations, twin boundaries and point defects.
 
More deformed crystals are difficult to simulate, both because models of the displacement field in such crystals are not easily obtainable and because the large strain would require impractical small step-sizes to ensure proper sampling of the scattering function. In these samples, dynamical diffraction effects are not thought to be important and the speckle-like noise due to lens aberrations should also be less strong, as the scattering is more diffuse. So a wave front based simulation approach is less appropriate in this type of samples. It may however be interesting to investigate the transition from the dynamical patterns from near-perfect crystallites to kinematical scattering from deformed crystals using our new approach.

\subsection*{Acknowledgements}

The authors kindly acknowledge the use of beamtime from the European Synchrotorn. M.C. and H.S. acknowledge funding from ERC Starting Grant \#804665.

\bibliography{large}{}
\bibliographystyle{unsrt}

\end{document}